\documentclass[12pt,onecolumn]{article}
\usepackage[dvips]{graphics}
\usepackage[dvips]{color}
\usepackage{amssymb}
\usepackage{amsmath}
\usepackage{amsthm}
\usepackage{url}

\theoremstyle{plain}
\newtheorem{theorem}{Theorem}
\newtheorem{lemma}{Lemma}

\def\tr{\mathop{\rm tr}\nolimits}%
\def\var{\mathop{\rm Var}\nolimits}%
\def\essinf{\mathop{\rm ess\,inf}}%
\def\sgn{\mathop{\rm sgn}\nolimits}%
\def\erfc{\mathop{\rm erfc}\nolimits}%
\def\inff{\mathop{\rm \vphantom{p}inf}}%

\newcommand{\iid}{\text{ i.i.d. }}

\newcommand{\geql}{\succeq}

\newcommand{\Real}{\mathbb{R}}
\newcommand{\eit}{e^{i\theta}}
\newcommand{\dth}{\,\frac{d\theta}{2\pi}}
\newcommand{\intt}{\int_{-\pi}^{\pi}}
\newcommand{\half}{\frac{1}{2}}

\begin{document}
\renewcommand{\textfraction}{0}

\title{On the Feedback Capacity of Stationary Gaussian Channels%
\footnote{This research is supported in part by NSF Grant CCR-0311633.}}
\author{\normalsize  
Young-Han Kim \\           
\small Information Systems Laboratory \\[-5pt] 
\small Stanford University \\[-5pt]
\small Stanford, CA 94305-9510 \\[-5pt] \small                 
yhk@stanford.edu}
\date{}
\maketitle
\thispagestyle{empty}

\begin{abstract}
The capacity of stationary additive Gaussian noise channels with
feedback is characterized as the solution to a variational problem.
Toward this end, it is proved that the optimal feedback coding scheme
is stationary.  When specialized to the first-order autoregressive
moving-average noise spectrum, this variational characterization
yields a closed-form expression for the feedback capacity.  In
particular, this result shows that the celebrated
Schalkwijk--Kailath coding scheme achieves the feedback
capacity for the first-order autoregressive moving-average Gaussian
channel, resolving a long-standing open problem studied by Butman,
Schalkwijk--Tiernan, Wolfowitz, Ozarow, Ordentlich,
Yang--Kav\v{c}i\'{c}--Tatikonda, and others.
\end{abstract}
\normalsize

\section{Introduction and summary}
\label{sec:intro}
We consider the additive Gaussian noise channel $Y_i = X_i + Z_i,$
$i=1,2,\ldots,$ where the additive Gaussian noise process
$\{Z_i\}_{i=1}^\infty$ is stationary with $Z^n = (Z_1,\ldots,Z_n) \sim
N_n(0, K_{Z,n})$ for each $n = 1,2,\ldots.$ We wish to communicate
a message $W \in \{1,\ldots,2^{nR}\}$ over the channel $Y^n = X^n +
Z^n$.  For block length $n$, we specify a $(2^{nR}, n)$ feedback code
with codewords $X^n(W,Y^{n-1}) = (X_1(W), X_2(W,Y_1), \ldots,
X_n(W,Y^{n-1})), W=1,\ldots,2^{nR},$ satisfying the average power
constraint
$
\frac{1}{n} \sum_{i=1}^n E X_i^2(W,Y^{i-1}) \le P
$
and decoding function $\hat{W}_n: \Real^n \to \{1,\ldots,2^{nR}\}.$
The probability of error $P_e^{(n)}$ is defined by
$ P_e^{(n)} = \Pr\{\hat{W}_n(Y^n) \ne W\}, $
where the message $W$ is uniformly distributed over
$\{1,2,\ldots,2^{nR}\}$ and is independent of $Z^n$.  We say that the
rate $R$ is achievable if there exists a sequence of $(2^{nR}, n)$
codes with $P_e^{(n)} \to 0$ as $n\to\infty$.  The feedback capacity
$C_\text{FB}$ is defined as the supremum of all achievable rates.  We
also consider the case in which there is no feedback, corresponding to
the codewords $X^n(W) = (X_1(W), \ldots, X_n(W))$ independent of the
previous channel outputs.  We define the nonfeedback capacity $C$, or
the capacity in short, in a manner similar to the feedback case.

Shannon~\cite{Shannon 1949} showed that the nonfeedback capacity is
achieved by water-filling on the noise spectrum, which is arguably one
of the most beautiful results in information theory.  More
specifically, the capacity $C$ of the additive Gaussian noise channel
$Y_i = X_i + Z_i,$ $i=1,2,\ldots,$ under the power constraint $P$, is
given by
\begin{equation}
\label{eq:c}
C = \intt \half \log
\frac{\max\{S_Z(\eit), \lambda\}}{S_Z(\eit)}\dth
\end{equation}
where $S_Z(\eit)$ is the power spectral density of the stationary
noise process $\{Z_i\}_{i=1}^\infty$ and the water-level $\lambda$ is
chosen to satisfy
\begin{equation}
\label{eq:p}
P = \intt \max\{0, \lambda -
S_Z(\eit)\}\dth.
\end{equation}
Although \eqref{eq:c} and \eqref{eq:p} give only a parametric
characterization of the capacity $C(\lambda)$ under the power
constraint $P(\lambda)$ for each parameter $\lambda \ge 0$, this
solution is considered to be simple and elegant enough to be called
\emph{closed-form}.

For the case of feedback, no such elegant solution exists.  Most
notably, Cover and Pombra~\cite{CP 1989} characterized the $n$-block
feedback capacity $C_{\text{FB},n}$ for arbitrary time-varying
Gaussian channels via the asymptotic equipartition property (AEP) for
arbitrary nonstationary nonergodic Gaussian processes as
\begin{equation}
\label{eq:cnfb}
C_{\text{FB},n} = \max_{K_{V,n}, B_n} \half \log
\frac{\det(K_{V,n} +
(B_n+I)K_{Z,n}(B_n+I)')^{1/n}}{\det(K_{Z,n})^{1/n}}
\end{equation}
where the maximum is taken over all positive semidefinite matrices
$K_{V,n}$ and all strictly lower triangular matrices $B_n$ of sizes
$n\times n$ satisfying $\tr(K_{V,n} + B_n K_{Z,n} (B_n)') \le nP$.
Note that we can also recover the nonfeedback case by taking $B_n
\equiv 0$.  When specialized to stationary noise processes, the
Cover--Pombra characterization gives the feedback capacity as a
limiting expression
\begin{align}
\nonumber
C_\text{FB} &= \lim_{n\to\infty} C_{\text{FB}, n}\\
&= \lim_{n\to\infty} \max_{K_{V,n}, B_n} \half \log
\frac{\det(K_{V,n} +
(B_n+I)K_{Z,n}(B_n+I)')^{1/n}}{\det(K_{Z,n})^{1/n}}.
\label{eq:cfb}
\end{align}

Despite its generality, the Cover--Pombra formulation of the feedback
capacity falls short of what we can call a closed-form solution.  It
is very difficult, if not impossible, to obtain an analytic expression
for the optimal $(K_{V,n}^{\star}, B_n^{\star})$ in \eqref{eq:cnfb}
for each $n$.  Furthermore, the sequence of optimal
$\{K_{V,n}^{\star}, B_n^{\star}\}_{n=1}^\infty$ is not necessarily
consistent, that is, $(K_{V,n}^{\star}, B_n^{\star})$ is not
necessarily a subblock of $(K_{V,n+1}^{\star}, B_{n+1}^{\star})$.
Hence the characterization \eqref{eq:cnfb} in itself does not give
much hint on the structure of optimal $\{K_{V,n}^{\star},
B_n^{\star}\}_{n=1}^\infty$ achieving $C_{\text{FB},n}$, or more
importantly, its limiting behavior.

In this paper, we make one step forward by proving
\begin{theorem}
\label{thm:fb-capacity}
The feedback capacity $C_\text{FB}$ of the Gaussian channel $Y_i = X_i
+ Z_i,$ $i = 1,2,\ldots,$ under the power constraint $P$, is given by
\[
C_\text{FB} = \sup_{S_V(\eit), B(\eit)} 
\intt \frac{1}{2} \log
\frac{S_V(\eit) + |1+B(\eit)|^2 S_Z(\eit)}{S_Z(\eit)} \dth
\]
where $S_Z(\eit)$ is the power spectral density of the noise process
$\{Z_i\}_{i=1}^\infty$ and the supremum is taken over all power
spectral densities $S_V(\eit) \ge 0$ and strictly causal filters
$B(\eit) = \sum_{j=1}^\infty b_j e^{ij\theta}$ satisfying the power
constraint $\frac{1}{2\pi}\intt (S_V(\eit) + |B(\eit)|^2 S_Z(\eit))
\,d\theta \le P.$
\end{theorem}
\noindent Roughly speaking, this characterization shows the asymptotic
optimality of a stationary solution $(K_{V,n}, B_n)$ in
\eqref{eq:cnfb} and hence it can be viewed as a justification for
interchange of the order of limit and maximum in \eqref{eq:cfb}.

Since Theorem 1 gives a variational expression of the feedback
capacity, it remains to characterize the optimal $(S_V^{\star}(\eit),
B^{\star}(\eit))$.  In this paper, we provide a sufficient condition
for the optimal solution using elementary arguments.  This result,
when specialized to the first-order autoregressive (AR) noise spectrum
$S_Z(\eit) = 1/{|1 + \beta\eit|^2}, -1<\beta<1,$ yields a closed-form
solution for feedback capacity as $C_\text{FB} = - \log x_0,$ where
$x_0$ is the unique positive root of the fourth-order polynomial
\[
P\,x^2 = \frac{(1-x^2)}{(1+|\beta|x)^2}.
\]
This result positively answers the long-standing conjecture by
Butman~\cite{Butman 1969, Butman 1976},
Tiernan--Schalkwijk \cite{TS 1974, Tiernan 1976}, and
Wolfowitz~\cite{Wolfowitz 1975}.  In fact, we will obtain the feedback
capacity formula for the first-order autoregressive moving average
(ARMA) noise spectrum, generalizing the result in \cite{Kim 2005} and
confirming a recent conjecture by Yang, Kav\v{c}i\'{c}, and
Tatikonda~\cite{YKT 2004}.

The rest of the paper is organized as follows.  We prove Theorem 1 in
the next section.  In Section 3, we derive a sufficient condition for
the optimal $(S_V^\star(\eit), B^\star(\eit))$ and apply this result
to the first-order ARMA noise spectrum to obtain the closed-form
feedback capacity.  We also show that the Schalkwijk--Kailath--Butman
coding scheme \cite{SK 1966, Schalkwijk 1966, Butman 1969} achieves
the feedback capacity of the first-order ARMA Gaussian channel.

\section{Proof of Theorem 1}
\label{sec:feedback}

We start from the Cover-Pombra formulation of the $n$-block feedback
capacity $C_{\text{FB},n}$ in
\eqref{eq:cnfb}.
Tracing the development of Cover and Pombra~\cite{CP 1989}
backwards, we express $C_{\text{FB},n}$ as
\begin{align}
\nonumber
C_{\text{FB},n} 
= \max_{V^n + B_n Z^n} h(Y^n) - h(Z^n)
= \max_{V^n + B_n Z^n} I(V^n;Y^n)
\end{align}
where the maximization is over all $X^n$ of the form $X^n = V^n + B_n
Z^n,$ resulting in $Y^n = V^n + (I+B_n)Z^n$, with strictly
lower-triangular $B_n$ and multivariate Gaussian $V^n,$ independent of
$Z^n$, satisfying the power constraint $E\sum_{i=1}^n X_i^2 \le nP.$

Define
\[
\tilde{C}_\text{FB} = \sup_{S_V(\eit), B(\eit)} 
\intt \frac{1}{2} \log
\frac{S_V(\eit) + |1+B(\eit)|^2 S_Z(\eit)} {S_Z(\eit)} \dth
\]
where $S_Z(\eit)$ is the power spectral density of the noise process
$\{Z_i\}_{i=1}^\infty$ and the supremum is taken over all power
spectral densities $S_V(\eit) \ge 0$ and strictly causal filters
$B(\eit) = \sum_{k=1}^\infty b_k e^{ik\theta}$ satisfying the power
constraint $\intt (S_V(\eit) + |B(\eit)|^2 S_Z(\eit)) \,d\theta \le
2\pi P.$ In the light of Szeg\H{o}--Kolmogorov--Krein theorem, we can
express $\tilde{C}_\text{FB}$ also as
\begin{equation}
\tilde{C}_{\text{FB}} = \sup_{\{X_i\}} h(\mathcal{Y}) - h(\mathcal{Z})
\nonumber
\end{equation}
where the supremum is taken over all stationary Gaussian processes
$\{X_i\}_{i=-\infty}^\infty$ of the form $X_i = V_i +
\sum_{k=1}^\infty b_k Z_{i-k}$ where $\{V_i\}_{i=-\infty}^\infty$ is
stationary and independent of $\{Z_i\}_{i=-\infty}^\infty$ such that
$E X_0^2 \le P$.  We will prove that $C_\text{FB} = \tilde{C}_\text{FB}$.

We first show that 
$C_{\text{FB},n} \le \tilde{C}_\text{FB}$
for all $n$.  Fix $n$ and let $(K_{V,n}^{\star}, B_n^{\star})$ achieve
$C_{\text{FB},n}$.  Consider a process $\{V_i\}_{i=-\infty}^{\infty}$
that is independent of $\{Z_i\}_{i=-\infty}^\infty$ and blockwise
i.i.d.\@ with $V_{kn+1}^{(k+1)n} \sim N_n(0,K_{V,n}^\star),$ $k =
0,\pm 1, \pm 2, \ldots.$ Define a process $\{X_i\}_{i=-\infty}^\infty$
as $X_{kn+1}^{(k+1)n} = V_{kn+1}^{(k+1)n} + B_n^\star
Z_{kn+1}^{(k+1)n}$ for all $k$.  Similarly, let $Y_i = X_i + Z_i,$
$-\infty < i < \infty,$ be the corresponding output process through the
stationary Gaussian channel.  Note that $Y_{kn+1}^{(k+1)n} =
V_{kn+1}^{(k+1)n} + (I + B_n^\star) Z_{kn+1}^{(k+1)n}$ for all $k$.
For each $t = 0,1,\ldots,n-1$, define a process
$\{V_{i}(t)\}_{i=-\infty}^\infty$ as $V_{i}(t) =V_{t+i}$ for all $i$
and similarly define $\{X_i(t)\}_{i=-\infty}^\infty$,
$\{Y_i(t)\}_{i=-\infty}^\infty$, and $\{Z_i(t)\}_{i=-\infty}^\infty$.
Note that $Y_i(t) = X_i(t) + Z_i(t)$ for all $i$ and all $t =
0,1,\ldots,n-1$, but $X_1^n(t)$ is \emph{not} equal to $V_1^n(t) +
B_n^{\star} Z_1^n(t)$ in general.

From the independence of $V_1^n$ and $V_{n+1}^{2n}$, we can easily check that
\begin{align*}
2 C_{\text{FB},n} &= I(V_1^{n}; Y_1^{n}) + I(V_{n+1}^{2n}; Y_{n+1}^{2n})\\
&= h(V_1^n) + h(V_{n+1}^{2n}) - h(V_1^n|Y_1^n) -
h(V_{n+1}^{2n}|Y_{n+1}^{2n})\\ &\le h(V_1^{2n}) - h(V_1^{2n}|Y_1^{2n})\\
&= I(V_1^{2n}; Y_1^{2n})\\ &= h(Y_1^{2n}) - h(Z_1^{2n}).
\end{align*}
By repeating the same argument, we get
\[
C_{\text{FB},n} \le \frac{1}{kn} (h(V_1^{kn}) - h(Z_1^{kn})),
\]
for all $k$.  Hence, for all $m = 1, 2,\ldots,$ and each $t =
0,\ldots, n-1$, we have
\begin{align*}
C_{\text{FB},n} &\le \frac{1}{m} \big( h(Y_{1}^m(t)) - h(Z_1^m(t)) \big) +
\epsilon_m\\
&= \frac{1}{m} \big( h(Y_{1}^m(t)) - h(Z_1^m) \big) + \epsilon_m
\end{align*}
where $\epsilon_m$ absorbs the edge effect and vanishes uniformly in
$t$ as $m\to\infty$.

Now we introduce a random variable $T$ uniform on $\{0,1,\ldots,n-1\}$
and independent of $\{V_i,X_i,Y_i,Z_i\}_{i=-\infty}^\infty$.  It is easy
to check the followings:

{\hbox{\rlap{(I)}\@\hphantom{(III)} 
$\{V_i(T), X_i(T), Y_i(T), Z_i(T)\}_{i=-\infty}^\infty$ is stationary with
$Y_i(T) = X_i(T) + Z_i(T)$.}}

{\hbox{\rlap{(II)}\@\hphantom{(III)} 
$\{X_i(T)\}_{i=-\infty}^\infty$ satisfies the power constraint}
\[E X_{0}^2(T) = E[E(X_{0}^2(T)|T)] = \frac{1}{n} \tr (K_{V,n}^\star 
B_n^\star K_{Z,n}(B_n^\star)') \le P.\]}

\noindent(III) $\{V_i(T)\}_{i=-\infty}^\infty$ and $\{Z_i(T)\}_{i=-\infty}^\infty$
are orthogonal; i.e., $E V_{i}(T)Z_{j}(T) = 0$ for all $i,j$.

\noindent (IV)
Although there is no linear relationship between $\{X_i(T)\}$ and
$\{Z_i(T)\}$, $\{X_i(T)\}$ 
still depends on $\{Z_i(T)\}$ in a strictly
causal manner.  More precisely, for all $i \le j$,
\begin{align*}
E (X_{i}(T) Z_{j}(T) | Z_{-\infty}^{i-1}(T)) 
&= E \bigl(
E (X_{i}(T) Z_{j}(T) | Z_{-\infty}^{i-1}(T), T) | Z_{-\infty}^{i-1}(T) 
\bigr)\\
&= E \bigl(
E (X_{i}(T) | Z_{-\infty}^{i-1}(T), T) 
E (Z_{j}(T) | Z_{-\infty}^{i-1}(T), T) 
| Z_{-\infty}^{i-1}(T) 
\bigr)\\
&= E \bigl(
E (X_{i}(T) | Z_{-\infty}^{i-1}(T), T) 
E (Z_{j}(T) | Z_{-\infty}^{i-1}(T)) 
| Z_{-\infty}^{i-1}(T) 
\bigr)\\
&= 
E (X_{i}(T) | Z_{-\infty}^{i-1}(T)) 
E (Z_{j}(T) | Z_{-\infty}^{i-1}(T)),
\end{align*}
and for all $i$,
\[
\var (X_{i}(T)-V_{i}(T) | Z_{-\infty}^{i-1}(T))
= E \bigl(
\var (X_{i}(T)-V_{i}(T) | Z_{-\infty}^{i-1}(T), T) | Z_{-\infty}^{i-1}(T)
\bigr)
= 0.
\]

Finally, define $\{\tilde{V}_i, \tilde{X}_i, \tilde{Y}_i,
\tilde{Z}_i\}_{i=-\infty}^\infty$ to be a jointly Gaussian stationary
process with the same mean and autocorrelation as $\{V_i(T), X_i(T),
Y_i(T), Z_i(T)\}_{i=-\infty}^\infty$.  It is easy to check that
$\{\tilde{V}_i, \tilde{X}_i, \tilde{Y}_i, \tilde{Z}_i\}$ also
satisfies the properties (I)--(IV) and hence that $\{\tilde{V}_i\}$
and $\{\tilde{Z}_i\}$ are independent.  It follows from these
properties and the Gaussianity of $\{\tilde{V}_i, \tilde{X}_i,
\tilde{Y}_i, \tilde{Z}_i\}$ that there exists a sequence
$\{b_k\}_{k=1}^\infty$ such that $\tilde{X}_i = \tilde{V}_i +
\sum_{k=1}^\infty b_k\tilde{Z}_{i-k}.$ Thus we have
\begin{align*}
C_{\text{FB},n} 
&\le \frac{1}{m} \bigl( h(Y_{1}^m(T)|T) - h(Z_1^m) \bigr) + \epsilon_m\\
&\le \frac{1}{m} \bigl( h(Y_{1}^m(T)) - h(Z_1^m) \bigr) + \epsilon_m\\
&\le \frac{1}{m} \bigl( h(\tilde{Y}_{1}^m) 
                      - h({Z}_1^m) \bigr) + \epsilon_m.
\end{align*}
By letting $m\to\infty$ and using the definition of
$\tilde{C}_\text{FB}$, we obtain
\begin{align}
\label{eq:c-n-tilde-ineq}
C_{\text{FB},n} \le h(\mathcal{\tilde{Y}}) - h(\mathcal{{Z}}) \le
\tilde{C}_\text{FB}.
\end{align}

For the other direction of the inequality, we use the notation
$\tilde{C}_\text{FB}(P)$ and $C_{\text{FB},n}(P)$ to stress the
dependence on the power constraint $P$.  Given $\epsilon > 0$, let
$\{{X}_i = {V}_i + \sum_{k=1}^\infty b_k
Z_{i-k}\}_{i=-\infty}^\infty$ achieve $\tilde{C}_\text{FB}(P) -
\epsilon$ under the power constraint $P$. The corresponding channel
output is given as
\begin{equation}
\label{eq:entropy-y}
{Y}_i = {V}_i + Z_i + \sum_{k=1}^\infty b_k Z_{i-k}
\end{equation}
for all $i = 0, \pm 1, \pm 2,\ldots.$

Now, for each $m = 1,2,\ldots,$ we define a single-sided nonstationary
process $\{X_i(m)\}_{i=1}^\infty$ in the following way:
\begin{align*}
X_i(m) &= 
\left\{
\begin{array}{ll}
U_i + V_i + \sum_{k=1}^{i-1} b_k Z_{i-k}&\qquad i \le m,\\
U_i + V_i + \sum_{k=1}^{m} b_k Z_{i-k}&\qquad i > m
\end{array}\right.
\end{align*}
where $U_1,U_2,\ldots$ are \iid $\sim N(0,\epsilon)$.  Thus, $X_i(m)$
depends causally on $Z_1^{i-1}$ for all $i$ and $m$. Let
$\{Y_i(m)\}_{i=1}^\infty$ be the corresponding channel output $Y_i(m) =
X_i(m) + Z_i,$ $i=1,2,\ldots,$ for each $m = 1,2,\ldots.$ We can show
that there exists an $m^{\star}$ so that
\begin{equation}
\nonumber
\lim_{n\to\infty} \frac{1}{n} \sum_{i=1}^n E X_i^2(m^{\star}) \le P + 2\epsilon
\end{equation}
and
\begin{equation}
\label{eq:ent-conv}
\lim_{n\to\infty} \frac{1}{n} h(Y_1^n(m^{\star}))
\ge h(\mathcal{Y}) - \epsilon
\end{equation}
where $h(\mathcal{Y})$ is the entropy rate of the stationary process
defined in~\eqref{eq:entropy-y}.  Consequently, for $n$ sufficiently
large,
\[
\frac{1}{n} \sum_{i=1}^n E X_i^2(m^{\star}) \le  n (P + 3\epsilon)
\]
and
\[
\frac{1}{n} (h(Y_1^n(m^{\star})) - h(Z_1^n)) \ge \tilde{C}_\text{FB}(P) - 2\epsilon.
\]
Therefore, we can conclude that
\[
C_{\text{FB},n}(P+3\epsilon) \ge \tilde{C}_\text{FB}(P) - 2\epsilon
\]
for $n$ sufficiently large.  Finally, using continuity of
$C_{\text{FB}}(P) = \lim_{n\to\infty} C_{\text{FB},n}(P)$ in $P$, we
let $\epsilon \to 0$ to get $ C_{\text{FB}}(P) \ge
\tilde{C}_\text{FB},$ which, combined with \eqref{eq:c-n-tilde-ineq},
implies that
\[
C_{\text{FB}}(P) = \tilde{C}_\text{FB}(P).
\]

\section{Example: First-order ARMA noise spectrum}
\label{sec:optimal}
With the ultimate goal of an explicit characterization of
$C_\text{FB}$ as a function of $S_Z$ and $P$, we wish to solve the
optimization problem
\begin{align}
\label{eq:var-opt1}
\begin{array}{l@{\quad}l}
\text{maximize}&\intt \log \bigl(S_V(\eit) 
+ |1+B(\eit)|^2 S_Z(\eit)\bigr) \dth\\[.25em]
\text{subject to}&B(\eit) \text{ strictly causal}\\[.25em]
&S_V(\eit) \ge 0\\[.25em]
&\intt S_V(\eit) + |B(\eit)|^2 S_Z(\eit) \dth \le P.
\end{array}
\end{align}
Suppose that $S_Z(\eit)$ is bounded away from zero.  Then, under the
change of variable
\[
S_Y(\eit) = S_V(\eit) +|1+B(\eit)|^2S_Z(\eit),
\]
we rewrite \eqref{eq:var-opt1} as
\begin{align}
\label{eq:var-opt2}
\begin{array}{l@{\quad}l}
\text{maximize}&{\intt \log S_Y(\eit) \dth}\\[0.25em]
\text{subject to}&B(\eit) \text{ strictly causal}\\[.25em]
&{S_Y(\eit) \ge |1+B(\eit)|^2 S_Z(\eit)}\\[.25em]
&{\intt S_Y(\eit) - (B(\eit) + B(e^{-i\theta}) + 1)S_Z(\eit) \dth \le P.}
\end{array}
\end{align}
Take any $\nu > 0$, $\phi, \psi_1 \in L_\infty,$ and
$\psi_2, \psi_3 \in L_1$ such that
$\phi(\eit) > 0,$
$\log \phi \in L_1,$
$\psi_1(\eit) = \nu - \phi(\eit) \ge 0,$
\[
\left[\hspace{0.3em}
\begin{matrix}
\psi_1(\eit) &\psi_2(\eit)\\[0.25em]
\overline{\psi_2(\eit)} & \psi_3(\eit)
\end{matrix}
\hspace{0.3em}\right] \geql 0,
\]
and $A(\eit) := \psi_2(\eit) + \nu S_Z(\eit) \in L_1$ is anticausal.
Since any feasible $B(\eit)$ and $S_Y(\eit)$ satisfy
\[
\left[\hspace{0.3em}
\begin{matrix}
S_Y(\eit) &1+B(\eit)\\[.25em]
1+\overline{B(\eit)} & S_Z^{-1}(\eit)
\end{matrix}
\hspace{0.3em}\right] \geql 0,
\]
we have
\[
\tr\left(
\left[\hspace{0.3em}
\begin{matrix}
S_Y&1+B\\[0.25em]
1+\overline{B}&S_Z^{-1}
\end{matrix}
\hspace{0.3em}\right]
\left[\hspace{0.3em}
\begin{matrix}
\psi_1 &\psi_2\\[0.25em]
\overline{\psi_2} & \psi_3
\end{matrix}
\hspace{0.3em}\right] \right) = \phi_1 S_Y + \psi_2 (1+\overline{B}) +
\overline{\psi_2}(1 + B) + \psi_3 S_Z^{-1} \ge 0.
\]
From the fact that $\log x \le x - 1$ for all $x \ge 0$, we get the inequality
\begin{align}
\nonumber
\log S_Y &\le -\log \phi + \phi S_Y - 1\\
\nonumber
&= -\log \phi + \nu S_Y - \psi_1 S_Y - 1\\
\label{eq:fb-var-ineq}
&\le -\log \phi + \nu S_Y + \psi_2(1+\overline{B}) 
+ \overline{\psi_2}(1+B) + \psi_3 S_Z^{-1} -1.
\end{align}
Further, since $A \in L_1$ is anticausal and $B \in H_\infty$ is
strictly causal, $A\overline{B} \in L_1$ is strictly anticausal and
\begin{equation}
\label{eq:fb-causal-int}
\intt A(\eit)\overline{B(\eit)} \dth = \intt
\overline{A(\eit)}B(\eit)\dth = 0.
\end{equation}
By integrating both sides of \eqref{eq:fb-var-ineq}, we get
\begin{align}
\nonumber
\intt \log S_Y &\le \intt - \log \phi +
\nu S_Y + \psi_2(1+\overline{B}) + \overline{\psi_2}(1+B) + \psi_3 S_Z^{-1} - 1\\
\nonumber
&\le \intt - \log \phi +
\nu \bigl((B\!+\!\overline{B}\!+\!1) S_Z + P\bigr)
+ \psi_2(1+\overline{B}) + \overline{\psi_2}(1+B) + \psi_3 S_Z^{-1} - 1\\
\nonumber
&=  \intt - \log \phi +
\psi_2 + \overline{\psi_2} + \psi_3 S_Z^{-1} + \nu (S_Z + P) -1
+ A\overline{B} + \overline{A}B \\
\label{eq:fb-var-ubd}
&= \intt -\log \phi + \psi_2 + \overline{\psi_2} + \psi_3 S_Z^{-1} + \nu(S_Z +  P) -1
\end{align}
where the second inequality follows from the power constraint
in~\eqref{eq:var-opt2} and the last equality follows from
\eqref{eq:fb-causal-int}.

Checking the equality conditions in \eqref{eq:fb-var-ubd}, we find the following
sufficient condition for the optimality of a specific $(S_Y(\eit),
B(\eit))$.
\begin{lemma}
\label{lemma:fb-var-suff}
Suppose $S_Z(\eit)$ is bounded away from zero.  Suppose $B(\eit) \in
H_\infty$ is strictly causal with
\begin{equation}
\label{eq:fb-var-power-eq}
\intt |B(\eit)|^2 S_Z(\eit) \dth = P.
\end{equation}
If there exists $\lambda > 0$ such that
\[
\lambda \le \essinf_{\theta\in[-\pi,\pi)} {|1+B(\eit)|^2 S_Z(\eit)}
\]
and that
\[
\frac{\lambda}{1+B(e^{-i\theta})} - B(\eit)S_Z(\eit) \in L_1
\]
is anticausal, then $B(\eit)$ along with $S_V(\eit) \equiv 0$ attains
the feedback capacity.
\end{lemma}

Now we turn our attention to the first-order autoregressive moving
average noise spectrum $S_Z(z)$, defined by
\begin{equation}
\label{eq:arma1-spectrum}
S_Z(\eit) = \left|\frac{1 + \alpha \eit}{1 + \beta \eit}\right|^2,
\qquad \alpha \in [-1, 1],\; \beta \in (-1, 1).
\end{equation}
This spectral density corresponds to the stationary noise process
defined by
$Z_i + \beta Z_{i-1}  =  U_i + \alpha U_{i-1},$
where $\{U_i\}_{i=-\infty}^\infty$ is a white Gaussian process with
zero mean and unit variance.  We find the feedback capacity 
of the first-order ARMA Gaussian channel in the following.
\begin{theorem}
\label{thm:arma1}
Suppose the noise process $\{Z_i\}_{i=1}^\infty$ has the power
spectral density $S_Z(z)$ defined in \eqref{eq:arma1-spectrum}.  Then,
the feedback capacity $C_\text{FB}$ of the Gaussian channel $Y_i = X_i
+ Z_i, \enspace i=1,2,\ldots,$ under the power constraint $P$, is given by
\[
C_\text{FB} = - \log x_0
\]
where $x_0$ is the unique positive root of the fourth-order polynomial
\begin{equation}
\label{eq:arma1-poly}
P\,x^2 = 
\frac{(1 - x^2)( 1 + \sigma \alpha x)^2}{(1 + \sigma \beta x)^2}
\end{equation}
and 
\[
\sigma = \sgn(\beta-\alpha)
= \left\{ 
\begin{array}{ll}
1, &\quad \beta \ge \alpha,\\
-1, &\quad \beta < \alpha.
\end{array}\right.
\]
\end{theorem}

\begin{proof}[\textnormal{\textbf{Proof sketch.}}]
Without loss of generality, we assume that $|\alpha| < 1$.  The case
$|\alpha| = 1$ can be handled by a simple perturbation argument.  When
$|\alpha| < 1$, $S_Z(\eit)$ is bounded away from zero, so that we can
apply Lemma~\ref{lemma:fb-var-suff}.

Here is the bare-bone summary of the proof:  We will take the feedback
filter of the form
\begin{equation}
\label{eq:b_z}
B(z) = \frac{1+\beta z}{1+\alpha z} \cdot \frac{yz}{1-\sigma xz}
\end{equation}
where $x \in (0,1)$ is an arbitrary parameter corresponding to each
power constraint $P \in (0,\infty)$ under the the choice of $y =
\frac{x^2-1}{\sigma x}\cdot\frac{1+\sigma \alpha x}{1+\sigma \beta x}.$
Then, we can show that
$B(z)$ satisfies the sufficient condition
in Lemma~\ref{lemma:fb-var-suff} under the power constraint
\[
P = \intt |B(\eit)|^2 S_Z(\eit) \dth
= \intt \frac{y^2}{|1-x\eit|^2} \dth
= \frac{y^2}{1-x^2}.
\]
The rest of the proof is the actual implementation of this idea.  We
skip the details.
\end{proof}

Although the variational formulation of the feedback capacity
(Theorem~\ref{thm:fb-capacity}), along with the sufficient condition
for the optimal solution (Lemma~\ref{lemma:fb-var-suff}), leads to the
simple closed-form expression for the ARMA(1) feedback capacity
(Theorem~\ref{thm:arma1}), one might be still left with somewhat
uneasy feeling, due mostly to the algebraic and indirect nature of the
proof.  Now we take a more constructive approach and interpret the
properties of the optimal feedback filter $B^\star$.

Consider the following coding scheme.  Let $V \sim N(0,1)$.  Over the
channel $Y_i = X_i + Z_i$, $i=1,2,\ldots,$ the transmitter initially
sends $X_1 = V$ and subsequently refines the receiver's knowledge by
sending
\begin{equation}
\label{eq:xn}
X_n = (\sigma x)^{-(n-1)} (V - \hat{V}_{n-1})
\end{equation}
where $x$ is the unique positive root of \eqref{eq:arma1-poly} and
$\hat{V}_{n} = E(V|Y_1,\ldots,Y_{n})$ is the minimum mean-squared
error estimate of $V$ given the channel output up to time $n$.  We
will show that
\[
\liminf_{n\to\infty} \frac{1}{n} I(V; \hat{V}_{n}) \ge \half \log
\left(\frac{1}{x^2}\right)
\]
while 
\[
\limsup_{n\to\infty} \frac{1}{n} \sum_{i=1}^n X_i^2 \le P,
\]
which proves that the proposed coding scheme achieves the feedback
capacity.

Define, for $n \ge 2$,
\[
Y'_n = d_n V + U_n + (-\alpha)^{n-1} (\alpha U_0 - \beta Z_0)
\]
where
\begin{align*}
d_n &= \left(\frac{1+\sigma \beta x}{1+\sigma \alpha x}\right)
( 1 - (-\sigma \alpha x)^n) (\sigma x)^{-(n-1)}.
\end{align*} 
Then one can show that $Y'_n$ can be represented as a linear combination
of $Y_1,\ldots, Y_n$ and hence that
\[
E (V - \hat{V}_n)^2 \le
E \left(V - \left(\sum_{k=2}^n d_k Y'_k\right)\right)^2.
\]
Furthermore, we can check that
\[
\frac{E (V - (\sum_{k=2}^{n} d_k Y'_k))^2}
{E (V - (\sum_{k=2}^{n-1} d_k Y'_k))^2} \to \frac{1}{x^2},
\]
whence
\[
\limsup_{n\to\infty} \frac{1}{n} \log E (V - \hat{V}_{n})^2 \le \log
\left(\frac{1}{x^2}\right)
\]
or equivalently,
\[
\liminf_{n\to\infty} \frac{1}{n} I(V; \hat{V}_{n}) \ge \half \log
\left(\frac{1}{x^2}\right).
\]

On the other hand, for $n \ge 2$,
\begin{align*}
E X_n^2 &= x^{-2(n-1)} E (V - \hat{V}_{n-1})^2
\le x^{-2(n-1)}{E \left(V - \left(\sum_{k=2}^{n-1} d_k Y'_k\right)\right)^2},
\end{align*}
which converges to
\begin{align*}
\lim_{n\to\infty} \frac{x^{-2(n-1)}}{\sum_{k=2}^{n-1} d_k^2}
= \frac{(1+\sigma \alpha x)^2}{(1 + \sigma \beta x)^2} \cdot (x^{-2} - 1)
= P.
\end{align*}
Hence, we have shown that the simple linear coding scheme
\eqref{eq:xn} achieves the ARMA(1) feedback capacity.

The coding scheme described above uses the minimum mean-square error
decoding of the message $V$, or equivalently, the joint typicality
decoding of the Gaussian random codeword $V$, based on the general
asymptotic equipartition property of Gaussian processes shown by Cover
and Pombra~\cite[Theorem 5]{CP 1989}.
Instead of the Gaussian codebook $V$, the transmitter initially sends
a real number $\theta$ which is chosen from some equally spaced signal
constellation $\Theta$, say, $ \Theta = \{-1, -1\!+\!\delta, \ldots,
1\!-\!\delta, 1\},$ $\delta = 2/(2^{nR} - 1),$ and subsequently
corrects the receiver's estimation error by sending $\theta -
\hat{\theta}_n$ (up to appropriate scaling as before) at time $n$, where
$\hat{\theta}_n$ is the \emph{minimum variance unbiased linear}
estimate of $\theta$ given $Y^{n-1}$.  Now we can verify that the
optimal \emph{maximum-likelihood} decoding is equivalent to finding
$\theta^* \in\Theta$ that is closest to $\hat{\theta}_n$, which
results in the error probability
\[
P_e^{(n)} \le \erfc\left(\sqrt{c_0 x^{-2n} / 2^{2nR}}\right)
\]
where $\erfc(x) = \frac{2}{\sqrt{\pi}} \int_{x}^\infty \exp(-t^2) dt$
is the complementary error function and $c_0$ is a constant
independent of $n$.  This proves that the Schalkwijk--Kailath--Butman
coding scheme achieves $C_\text{FB} = -\log x$ with doubly
exponentially decaying error probability.

\section{Concluding remarks}
Although it is still short of what we can call a closed-form solution
in general, our variational characterization of Gaussian feedback
capacity gives an exact analytic answer for a certain class of
channels, as demonstrated in the example of the first-order ARMA
Gaussian channel.  Our development can be further extended in two
directions.  First, one can investigate properties of the optimal
solution $(S_V^\star, B^\star)$.  Without much surprise, one can show
that feedback increases the capacity if and only if the noise spectrum
is white.  Furthermore, it can be shown that taking $S_V^\star \equiv
0$ does not incur any loss in maximizing the output entropy, resulting
in a simpler maximin characterization of feedback capacity:
\[
C_{\text{FB}} = \sup_{\{b_k\}} \inff_{\{a_k\}}
\half \log \left(
\intt \Big|1-\sum_{k=1}^\infty a_k e^{ik\theta}\Big|^2
\Big|1-\sum_{k=1}^\infty b_k e^{ik\theta}\Big|^2
S_Z(\eit) \dth\right)
\]
\vskip -6pt
\noindent where the supremum is taken over all $\{b_k\}$
satisfying $ \intt \Big|\sum_{k=1}^\infty b_k e^{ik\theta}
\Big|^2 S_Z(\eit) \dth
\le P.$
  
Secondly, one can focus on the finite-order ARMA noise spectrum and
show that the $k$-dimensional generalization of
Schalkwijk--Kailath--Butman coding scheme is optimal for the ARMA
spectrum of order $k$.  This confirms many conjectures based on
numerical evidences, including the recent study by Yang,
Kav\v{c}i\'{c}, and Tatikonda~\cite{YKT 2004}.  These results will be
reported separately in \cite{Kim 2005b}.

\end{document}